\documentclass[twocolumn]{jpsj3}

\usepackage{dcolumn}
\usepackage{bm}
\usepackage{color}
\usepackage{txfonts}
\def\vector#1{{\boldsymbol{#1}}}

\def\vc{{\vector c}}
\def\vd{{\vector d}}
\def\vg{{\vector g}}

\def\vH{{\vector H}}
\def\vk{{\vector k}}

\def\vz{{\vector z}}

\def\Tc{{T_{\rm c}}}

\def\dps{\displaystyle}

\def\kB{{k_{\rm B}}}

\def\eq.#1{Eq.~(\ref{#1})}
\def\eqs.#1{Eqs.~(\ref{#1})}

\def\Hc2{{H_{\rm c2}}}
\def\difHc2{{H'_{\rm c2}}}

\newcommand\Equation[2]{
\begin{equation}\label{#1} 
#2
\end{equation}
}

\title{
Temperature Dependence of the Spin Susceptibility in \\
Noncentrosymmetric Superconductors with Line Nodes 
}

\author{Hiroshi Shimahara}

\inst{
Department of Quantum Matter Science, ADSM, Hiroshima University, 
Higashi-Hiroshima 739-8530, Japan
}

\recdate{December 4, 2013}

\abst{
The spin susceptibility of noncentrosymmetric superconductors is studied 
when the gap function has line nodes. 
As examples, d-wave states, 
where the gap function has an additional odd-parity phase factor, 
are examined. 
The curve of the spin susceptibility $\chi(T)$ is upward convex 
when all line nodes are parallel to the magnetic field, 
while it is downward convex 
in the other d-wave states and an s-wave state. 
For polycrystalline powder samples, 
the temperature dependences of $\chi(T)$ are predicted 
by assuming three explicit conditions of the powder particles. 
The results are compared with the experimental data of the Knight shift 
observed in ${\rm Li_2Pt_3B}$ and ${\rm Li_2Pd_3B}$.

}

\begin{document}
\sloppy
\maketitle

\newpage

\section{\label{sec:introduction}
Introduction 
}

Noncentrosymmetric superconductors 
${\rm Li_2Pt_3B}$ and ${\rm Li_2Pd_3B}$ 
exhibit quite different superconductivity behavior 
despite the similarity of their chemical and crystal structures. 
The superconducting transition temperature $\Tc$ is 7~K 
for ${\rm Li_2Pd_3B}$~\cite{Tog04}, 
while it is 2.7~K for ${\rm Li_2Pt_3B}$~\cite{Bad05}. 
Regarding the pairing anisotropy, 
most of the experimental results, 
such as the temperature dependences of 
the nuclear magnetic relaxation (NMR) rate $T_1^{-1}$~\cite{Nis05,Nis07}, 
magnetic penetration depth~\cite{Yua06}, and 
specific heat~\cite{Tak05,Tak07}, 
indicate that the gap function has no nodes in ${\rm Li_2Pd_3B}$, 
while it has line nodes in ${\rm Li_2Pt_3B}$, 
although the $H$--$T$ phase diagram 
for ${\rm Li_2(Pd_{1-x}Pt_x)_3B}$ 
is qualitatively unchanged for $0 \leq x \leq 1$~\cite{Pee11}. 
The temperature dependence of the Knight shift 
observed by Nishiyama et al.~\cite{Nis05,Nis07} 
indicates a full-gap state for ${\rm Li_2Pd_3B}$, 
which is consistent with the above experiments, 
while for ${\rm Li_2Pt_3B}$, 
the interpretation of the temperature dependence is nontrivial.

In ${\rm Li_2Pt_3B}$, 
the Knight shift remains unchanged 
below $\Tc$ 
within experimental resolution~\cite{Nis05,Nis07}, 
which indicates that the spin susceptibility is not reduced 
by the growth of the superconducting gap. 
Therefore, this implies that the superconducting state does not contain 
Cooper pairs of antiparallel-spin electrons, 
where the spin quantization axis is parallel to the magnetic field. 
However, such a superconducting state 
that consists purely of parallel-spin pairs 
cannot occur in systems with strong spin-orbit coupling, 
except in rare situations.

To discuss this issue, 
let us consider the bilinear terms of the Hamiltonian 
\Equation{eq:H0}
{
     H_0 = 
         \sum_{\vk} 
         \sum_{\sigma_1 \sigma_2} 
         c_{\vk \sigma_1}^{\dagger}
         \Bigl [ 
         \xi_{\vk}^0 \, \sigma_0 
           - \alpha_{\vk} \, {\hat {\vg}}(\vk) \cdot \boldsymbol{\sigma} 
         \Bigr ]_{\sigma_1 \sigma_2} 
         c_{\vk \sigma_2}, 
     }
where $\sigma_0$, $\boldsymbol{\sigma}$, and $c_{\vk \sigma}$ 
are the $2 \times 2$ identity matrix, 
the Pauli matrices, 
and the annihilation operator of the electron 
with momentum $\vk$ and spin $\sigma$, respectively. 
The vector function ${\hat \vg}(\vk)$ is assumed to satisfy 
${\hat \vg}(-\vk) = - {\hat \vg}(\vk)$ and \mbox{$|{\hat \vg}(\vk)| = 1$}. 
We introduce the polar coordinates 
$({\bar \theta}_{\vk},{\bar \varphi}_{\vk})$ 
for the direction of ${\hat \vg}(\vk)$ by 
\Equation{eq:gthetavarphidef}
{
     \begin{split}
     {\hat \vg}(\vk) 
         & = (g_x(\vk), g_y(\vk), g_z(\vk)) \\
         & = (\sin {\bar \theta}_{\vk} \cos {\bar \varphi}_{\vk}, 
         \sin {\bar \theta}_{\vk} \sin {\bar \varphi}_{\vk}, 
                            \cos {\bar \theta}_{\vk} ) , 
     \end{split}
     }
where the $z$-axis lies along the magnetic field direction. 
$H_0$ is diagonalized 
by a unitary transformation of the electron operators 
in spin space~\cite{Gor01,Ser04,Sam08,Shi13}, 
leading to the spin-orbit split bands having the one-particle energies 
$
     {\tilde \xi}_{\vk s} 
       = 
         \xi_{\vk}^0 - s \alpha_{\vk} $, 
where $s=\pm$. 
With $c_{\vk \sigma}$ and the annihilation operator ${\tilde c}_{\vk s}$ 
of the electron with momentum $\vk$ in the $s$-band, 
the superconducting order parameters are written as 
$\psi_{\sigma \sigma'} (\vk) 
   = \langle c_{\vk \sigma} c_{- \vk \sigma'} \rangle $ 
and 
${\tilde \psi}_{s s'} (\vk) 
   = \langle {\tilde c}_{\vk s} {\tilde c}_{- \vk s'} \rangle $. 
The former is expressed 
as 
$\psi_{\uparrow \uparrow} = - d_x +  i d_y$, 
$\psi_{\downarrow \downarrow} = d_x +  i d_y$, 
$\psi_{\uparrow \downarrow} = d_z + d_0$, 
and 
$\psi_{\downarrow \uparrow} = d_z - d_0$, 
in terms of 
the d-vector $\vd(\vk) = (d_x(\vk), d_y(\vk), d_z(\vk)) $ 
and the singlet component $d_0(\vk)$ 
of the order parameter.

When $\alpha_{\vk} \gg \kB \Tc$, 
the spin-orbit splitting of the Fermi-surfaces is so large that 
interband pairing does not occur, 
that is, 
${\tilde \psi}_{\pm \mp}(\vk) = 0$. 
This immediately leads to $\vd(\vk) \parallel {\hat {\vg}}(\vk)$ 
as Frigeri et al. discovered~\cite{Fri04a}. 
In this case, we can define a scalar function $d(\vk)$ by 
\Equation{eq:scalar_d_def}
{
     \vd(\vk) = d(\vk) {\hat {\vg}}(\vk) , 
     }
leading to 
\Equation{eq:psi_ss}
{
     {\tilde \psi}_{ss}(\vk) = s_{\vk} ( d(\vk) + s d_0(\vk) ) , 
     }
where $s_{\vk}$ is an odd-parity phase factor that originates from 
the unitary transformation~\cite{Gor01,Ser04,Shi13}. 
From the Knight shift data in ${\rm Li_2Pt_3B}$ mentioned above, 
if we assume that antiparallel-spin pairing is suppressed, 
$\psi_{\uparrow \downarrow}(\vk) = 
 \psi_{\downarrow \uparrow}(\vk) = 0$, 
i.e., 
$d_0(\vk) = d_z(\vk) = 0$, 
it follows that $d(\vk) = 0$ from \eq.{eq:scalar_d_def} 
unless ${\hat g}_z(\vk) = 0$. 
Hence, all components of the superconducting order parameter vanish, 
that is, $d_0(\vk) = 0$ and $\vd(\vk) = 0$. 
Therefore, pure parallel-spin pairing can occur over regions of $\vk$'s 
that satisfy the conditions 
${\bar \theta}_{\vk} \approx \pi/2$ 
or $\alpha_{\vk} \approx 0$.

It seems unusual that such a limited region has 
a sufficiently large density of states 
to yield the observed transition temperature. 
Even if this was possible in single crystal samples 
for an appropriate magnetic field direction, 
such a condition of the magnetic field direction 
is not satisfied in polycrystalline powder samples 
in which the orientation of each powder particle is random. 
Therefore, the interpretation of the Knight shift data 
for ${\rm Li_2Pt_3B}$ appears problematic. 
However, this difficulty can be resolved as we shall examine below, 
if we assume that the system is affected by the applied magnetic field.

In \eq.{eq:psi_ss}, 
$d(\vk)$ and $d_0(\vk)$ are of even parity from their definitions. 
Therefore, if parity-mixing terms are ignored in pairing interactions, 
the gap function is expanded as 
\Equation{eq:psiss_exp}
{
     \Delta_{\vk s} 
       = s_{\vk} \sum_{\alpha ({\rm even})} 
           \Delta_{\alpha}^{(s)}
             \gamma_{\alpha}^{(s)} (\vk) , 
     }
with basis functions $\gamma_{\alpha}^{(s)}(\vk)$, 
where $\alpha$ denotes the symmetry index 
and the summation is taken over $\alpha$'s of even parity. 
For example, 
$\alpha = (l,m)$ is convenient in spherically symmetric systems, 
where $l$ and $m$ are the quantum numbers of angular momentum. 
Since the phase factor $s_{\vk}$ has nothing to do with 
the quasi-particle energy 
$E_{\vk s} = \sqrt{{\tilde \xi}_{\vk s}^2 + |\Delta_{\vk s}|^2}$, 
it is appropriate to index the gap function by the value of $\alpha$ 
that is dominant in the summation in \eq.{eq:psiss_exp}. 
Therefore, the lowest-order line-node state is a d-wave~\cite{Shi13}.

An interesting problem to consider 
is how the difference in the superconductivity 
between ${\rm Li_2Pd_3B}$ and ${\rm Li_2Pt_3B}$ 
arises in spite of their similarity. 
This difference can be attributed to differences in the pairing interactions 
and the one-particle dispersion energy~\cite{Shi13}. 
In the case that those differences originate from differences 
in the strength of the spin-orbit interactions, 
a transition between a full-gap state and a line-node state would occur 
if we could continuously increase the spin-orbit coupling constant 
between the two compounds.

Shishidou and Oguchi have obtained spin-orbit split Fermi-surfaces 
for these compounds 
by first-principles calculations~\cite{Shis11}. 
Their results suggest that every Fermi surface has 
a spin-orbit split partner in ${\rm Li_2Pd_3B}$, 
while in ${\rm Li_2Pt_3B}$, 
many of the Fermi surfaces do not have partners 
because of stronger spin-orbit coupling.

In our previous work~\cite{Shi13}, 
we proposed a scenario in which 
the disappearance of one of the spin-orbit split Fermi surfaces 
in ${\rm Li_2Pt_3B}$ 
is mainly responsible for the difference observed in the superconductivity. 
Examining several types of pairing interactions, 
it was found that, 
when a charge--charge interaction is dominant, 
the transition from a full-gap state to a line-node state occurs 
over a wide and realistic region of the parameter space 
of the coupling constants for the interaction 
with increasing the spin-orbit coupling constant. 
If this scenario holds for the present compounds, 
presumably an s-wave nearly-spin-triplet state 
and a d-wave mixed-singlet-triplet state are realized 
in ${\rm Li_2Pd_3B}$ and ${\rm Li_2Pt_3B}$, respectively. 
These states are consistent with most of the available experimental data. 
For the Knight shifts in the superconducting states, 
the temperature dependence observed in ${\rm Li_2Pd_3B}$ 
can be understood by assuming an s-wave state, 
independently of the weights of the spin-singlet and triplet components, 
as shown below, 
while in ${\rm Li_2Pt_3B}$ it is nontrivial, as explained above.

In the present work, 
we examine the temperature dependence of the spin susceptibility 
for noncentrosymmetric superconductors, 
when the gap function has line nodes. 
We discuss a scenario in which 
the temperature dependences of the Knight shifts 
are consistently reproduced 
for ${\rm Li_2Pd_3B}$ and ${\rm Li_2Pt_3B}$, 
assuming s-wave and d-wave states, respectively, 
without specifying the microscopic origin of the pairing interactions. 
This assumption is phenomenologically plausible 
from the experimental results mentioned above, 
and consistent with the scenario proposed 
in our previous paper.~\cite{Shi13}

The spin susceptibility of noncentrosymmetric superconductors 
has been studied 
by many authors~\cite{Ede89,Ede95,Gor01,Yip02,Fri04b,
Fuj05,Fuj07b,Sam07,Muk12,Mar13}. 
In particular, it has been found that the spin susceptibility has 
a large Van Vleck component $\chi_{\rm V}$ 
that is almost temperature independent in the superconducting 
phase~\cite{Gor01,Yip02,Fri04b,Fuj05,Fuj07b,Sam07,Muk12,Mar13}.

The behavior of the Knight shift in ${\rm Li_2Pt_3B}$ implies that 
the difference in the spin susceptibilities 
$\Delta \chi \equiv \chi_{\rm N} - \chi_{\rm S}$ is small, 
where the subscripts N and S denote the normal and superconducting phases, 
respectively. 
Maruyama and Yanase obtained 
$\Delta \chi/\chi_{\rm N} = 1 - \chi_{\rm S}/\chi_{\rm N} < 0.1$, 
which is consistent with the experimental results for ${\rm Li_2Pt_3B}$, 
considering the reduction of the density of states 
for ${\rm Li_2Pt_3B}$ because of stronger spin-orbit coupling~\cite{Mar13}.

This small value of $\Delta \chi/\chi_{\rm N}$ is obtained by 
considering the large Van Vleck component $\chi_{\rm V}$, 
which significantly reduces the ratio $\Delta \chi/\chi_{\rm N}$. 
However, concerning the comparison of two compounds, 
the relevant quantity is 
the ratio $\Delta \chi^{\rm Pt}/\Delta \chi^{\rm Pd}$ 
rather than the ratio $\Delta \chi^{\rm Pt}/\chi_{\rm N}^{\rm Pt}$, 
where the superscripts Pt and Pd represent 
${\rm Li_2Pt_3B}$ and ${\rm Li_2Pd_3B}$ systems, respectively. 
The Van Vleck component $\chi_{\rm V}$ would not 
significantly change the ratio $\Delta \chi^{\rm Pt}/\Delta \chi^{\rm Pd}$ 
because it would reduce both $\Delta \chi^{\rm Pt}$ and $\Delta \chi^{\rm Pd}$ 
to a similar extent. 
Therefore, it seems that 
the smallness of $\Delta \chi^{\rm Pt}/\Delta \chi^{\rm Pd}$ 
observed by the Knight shift measurement is not completely explained 
only by the reduction of the density of states. 
Moreover, if $\Delta \chi^{\rm Pt}/\Delta \chi^{\rm Pd}$ is small 
merely because of the small density of states, 
$\Tc$ should be negligibly small in ${\rm Li_2Pt_3B}$ 
in comparison to that in ${\rm Li_2Pd_3B}$, 
unless the pairing interaction is extremely strong in ${\rm Li_2Pt_3B}$.

In Sect.~2, an expression for the spin susceptibility is presented. 
In Sect.~3, the spin susceptibilities are numerically calculated 
for various d-wave states using a simplified model. 
The results are compared with Knight shift data~\cite{Nis07} 
for ${\rm Li_2Pd_3B}$ and ${\rm Li_2Pt_3B}$. 
The final section summarizes the results.

\section{\label{sec:formulation}
Formulation 
}

We briefly review the expression for the spin susceptibility 
to clarify the notation. 
The total magnetization 
is expressed as 
${
     M = \mu_{\rm e} \langle {\hat m} \rangle 
     }$
with 
\Equation{eq:mhat_sigma}
{
     {\hat m} = 
     \sum_{i}
     \sum_{\sigma,\sigma'} 
     c_{i \sigma}^{\dagger} 
     \sigma_{\sigma \sigma'}^{z} 
     c_{i \sigma'} , 
     }
where the index $i$ denotes the lattice site, 
and $\mu_{\rm e}$ is the electron magnetic moment. 
The Zeeman energy term of the Hamiltonian is 
$ H_m = - \mu_{\rm e} H {\hat m} $, 
for a magnetic field $\vH = (0,0,H)$. 
The spin susceptibility per site is calculated by the formula: 
\Equation{eq:chiformula}
{
     \chi 
     = 
     i \int_{0}^{\infty} \hspace{-1ex} d t \,\, 
       \frac{1}{N} 
       \langle [ {\hat m}(t), {\hat m}(0) ] 
       \rangle . 
     }
In the superconducting state, the spin susceptibility is obtained as 
$ \chi = \chi_1 + \chi_2 + \chi_3$, where 
\Equation{eq:chi1}
{
     \chi_1 = \frac{1}{N} \sum_{\vk,s} 
                 \cos^2 {\bar \theta}_{\vk} \, 
                 \Bigl [ - \frac{d}{d E} f(E) 
                 \Bigr ]_{E = E_{\vk s}} , 
     }
\Equation{eq:chi2}
{
     \chi_2 = - \frac{2}{N} \sum_{\vk} \sin^2 {\bar \theta}_{\vk} \, 
        \Bigl [ n_{+-}(\vk) \Bigr ]^2 \, 
        \frac{f(E_{\vk +}) - f(E_{\vk -})}
             {  E_{\vk +}  -   E_{\vk -} } , 
     }
\Equation{eq:chi3}
{
     \chi_3 = - \frac{2}{N} \sum_{\vk} \sin^2 {\bar \theta}_{\vk} \, 
        \Bigl [ m_{+-}(\vk) \Bigr ]^2 \, 
        \frac{f(E_{\vk +}) - f(- E_{\vk -})}
             {  E_{\vk +}  +   E_{\vk -} } , 
     }
and 
$n_{+-}(\vk) = u_{\vk +} u_{\vk -} - v_{\vk +} v_{\vk -}$, 
$m_{+-}(\vk) = u_{\vk +} v_{\vk -} + v_{\vk +} u_{\vk -}$, 
$    u_{\vk s} = \bigl [ 
                 \bigl ( 1 + {\tilde \xi}_{\vk s}/E_{\vk s}
                 \bigr )/2
                 \bigr ]^{1/2}$, 
and 
$  v_{\vk s} = s \bigl [ 
                 \bigl ( 1 - {\tilde \xi}_{\vk s}/E_{\vk s}
                 \bigr )/2
                 \bigr ]^{1/2}$.

When $\alpha_{\vk} \gg |\Delta_{\vk s}|$, 
the temperature dependence of the spin susceptibility mainly 
occurs from $\chi_1$, 
and the interband component $\chi_2 + \chi_3 = \chi_{\rm V}$ 
barely depends on the temperature~\cite{Gor01,Yip02}. 
Therefore, the reduction of the spin susceptibility 
in the superconducting phase is 
$\Delta \chi \equiv \chi_{\rm N} - \chi_{\rm S}
             \approx \chi_{\rm 1N} - \chi_{\rm 1S}$. 
The difference in the Knight shifts 
between the superconducting and normal phases 
is expressed as $\Delta K= |A_{\rm hf}| \Delta \chi$ 
in terms of $\Delta \chi$ 
with the hyperfine coupling constant $A_{\rm hf} < 0$ 
between the nuclear and electron spins. 
Since $\chi_{\rm N}$ and $\chi_{\rm 1N}$ are almost constant for the metals, 
and $\chi_{\rm 1S} (0) = 0$, we obtain 
$\Delta \chi(0) 
   = \chi_{1{\rm N}} 
   = \langle \cos^2 {\bar \theta}_{\vk} \rangle_{\rm F} \chi_{\rm N}$, 
where $\langle \cdots \rangle_{\rm F}$ 
denotes the average over the Fermi surface. 
For example, in spherically symmetric systems, 
$\langle \cos^2 {\bar \theta}_{\vk} \rangle_{\rm F} = 1/3$ 
and $\Delta \chi(0) = \chi_{\rm N}/3$~\cite{Gor01,Yip02}. 
In planar systems in which 
${\hat \vg}(\vk) \perp {\hat \vz}$ for all $\vk$, 
$\langle \cos^2 {\bar \theta}_{\vk} \rangle_{\rm F} = 0$ 
and $\Delta \chi(0) = 0$, when $\vH \parallel {\hat \vz}$. 
In general, the ratio $\Delta \chi(0)/\chi_{\rm N} 
= \langle \cos^2 {\bar \theta}_{\vk} \rangle_{\rm F}$ 
is much smaller than the value 1 
for centrosymmetric singlet superconductors. 
However, considering the similarity of 
the ${\rm Li_2Pd_3B}$ and ${\rm Li_2Pt_3B}$ crystal structures, 
the averages $\langle \cos^2 {\bar \theta}_{\vk} \rangle_{\rm F}$ 
for the two compounds would be roughly canceled out 
in $\Delta \chi^{\rm Pt}(0)/\Delta \chi^{\rm Pd}(0)$ at $T = 0$.

At finite temperatures below $\Tc$, 
the temperature dependence of $\Delta \chi(T)$ differs qualitatively 
depending on the pairing anisotropy. 
As is well known, 
$\Delta \chi(T)$ is proportional to the Yosida function 
in the s-wave state, 
while it is proportional to $T$ at low temperatures 
in the line-node states. 
In addition to this difference, 
the factor $\cos^2 {\bar \theta}_{\vk}$ in $\chi_1$ 
gives rise to qualitatively different temperature dependences 
in the d-wave states for noncentrosymmetric superconductors. 
When the gap function has a peak near ${\bar \theta}_{\vk} = \pi/2$, 
the growth of the superconducting gap is less effective at reducing 
the susceptibility $\chi_1$. 
As a result, the difference $\Delta \chi = \chi_{\rm N} - \chi_{\rm S}$ 
becomes smaller. 
Similarly, when the gap function has a peak near 
${\bar \theta}_{\vk} = 0$ or $\pi$, 
the growth of the superconducting gap is more effective at reducing 
the susceptibility $\chi_1$, 
and the difference $\Delta \chi$ becomes larger.

To illustrate this phenomenon, 
we suppose a spherically symmetric system 
in which ${\hat \vg}(\vk) = \vk/|\vk| \equiv {\hat \vk}$. 
The basis functions are written as 
$$ {
     \gamma_{lm}^{(s)}(\vk) 
       = C_{lm}^{(s)} \, 
         \theta( \omega_{\rm c}^{(s)} - | {\tilde \xi}_{\vk s} | ) \, 
         Y_{lm}({\hat \vk}) , 
     }$$ 
in the weak coupling theory, 
where $C_{lm}^{(s)}$ and $\omega_{\rm c}^{(s)}$ 
are the normalization factor and 
the cutoff energy of the pairing interactions, respectively. 
Here, we have defined the spherical harmonic function by 
$
     Y_{lm}({\hat \vk}) 
     = P_{l}^{m}(\cos \theta_{\hat \vk}) e^{i m \varphi_{\hat \vk}} 
     $, 
where $\theta_{\hat \vk}$ and $\varphi_{\hat \vk}$ are 
the polar and azimuthal angles of the direction of ${\hat \vk}$, 
respectively. 
We examine 
${\rm d}_{xy}$, 
${\rm d}_{yz}$, 
${\rm d}_{zx}$, 
${\rm d}_{x^2-y^2}$, and 
${\rm d}_{3z^2-r^2}$ wave states 
as examples of the line-node state, 
and the s-wave state as the full-gap state. 
The gap functions for these d-wave states are 
\Equation{eq:Delta_d-wave}
{
     \begin{split}
     \Delta_{\vk +}^{(xy)}
     & = 
         (15/4)^{\frac{1}{2}}
         s_{\vk} 
         \Delta_{xy}
         \sin^2 {\theta}_{\hat \vk} 
         \sin 2 {\varphi}_{\hat \vk}  , 
     \\
     \Delta_{\vk +}^{(x^2-y^2)}
     & = 
         (15/4)^{\frac{1}{2}}
         s_{\vk} 
         \Delta_{x^2-y^2}
         \sin^2 {\theta}_{\hat \vk} 
         \cos 2 {\varphi}_{\hat \vk} , 
     \\
     \Delta_{\vk +}^{(yz)}
     & = 
         15^{\frac{1}{2}}
         s_{\vk} 
         \Delta_{yz}
         \sin {\theta}_{\hat \vk} \cos {\theta}_{\hat \vk} 
         \sin {\varphi}_{\hat \vk} , 
     \\
     \Delta_{\vk +}^{(zx)}
     & = 
         15^{\frac{1}{2}}
         s_{\vk} 
         \Delta_{zx}
         \sin {\theta}_{\hat \vk} \cos {\theta}_{\hat \vk} 
         \cos {\varphi}_{\hat \vk} , 
     \\
     \Delta_{\vk +}^{(3z^2-r^2)}
     & = 
         (5/4)^{\frac{1}{2}}
         s_{\vk} 
         \Delta_{3z^2-r^2}
         (3 \cos^2 {\theta}_{\hat \vk} - 1) , 
     \end{split}
     }
on the Fermi surface. 
For ${\rm Li_2Pt_3B}$, we assume that 
the Fermi surface vanishes in the band with $s = -$.

\begin{figure}
\vspace{4ex} 
\begin{center}
\includegraphics[width=5.5cm]{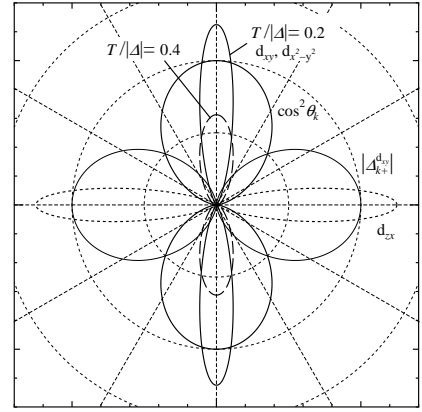}
\end{center}
\caption{
Angular dependence of 
$- (\Delta_{\vk +}/T) f'(\Delta_{\vk +})$ 
on the Fermi surface, 
where 
$\varphi_{\vk} = \pi/4$ 
for the ${\rm d}_{xy}$ and ${\rm d}_{yz}$ wave states 
and 
$\varphi_{\vk} = 0$ 
for the ${\rm d}_{x^2-y^2}$ wave state. 
} 
\label{fig:chi1integrand}
\end{figure}

Figure~\ref{fig:chi1integrand} plots the angular dependences 
of the factor 
$\cos^2 {\bar \theta}_{\vk} = \cos^2 \theta_{\hat \vk}$ 
and the function 
$- (\Delta_{\vk +}/T) f'(\Delta_{\vk +})$ 
on the Fermi surface 
that appear in the integral of $\chi_1$ in \eq.{eq:chi1}. 
For the ${\rm d}_{xy}$- and ${\rm d}_{x^2-y^2}$-wave states, 
the function $- (\Delta_{\vk +}/T) f'(\Delta_{\vk +})$ is large 
where the factor $\cos^2 \theta_{\vk}$ is large. 
Hence, in these states, 
$\chi_{1{\rm S}}/\chi_{1{\rm N}} = 1 - \Delta \chi_1/\chi_{1{\rm N}}$ 
turns out to be large 
and the difference $\Delta K = |A_{\rm hf}| \Delta \chi$ 
thus becomes small. 
In the ${\rm d}_{yz}$-wave state, however, 
the situation is contrary to this.

\section{\label{sec:results}
Numerical Results 
}

In this section, we calculate $\chi_1$ 
that contributes to the temperature-dependent component of the Knight shift. 
We solve the gap equation numerically~\cite{Shi13}. 
For simplicity, 
we do not consider mixing different d-wave order parameters.

\begin{figure}
\vspace{4ex} 
\begin{center}
\includegraphics[width=8cm]{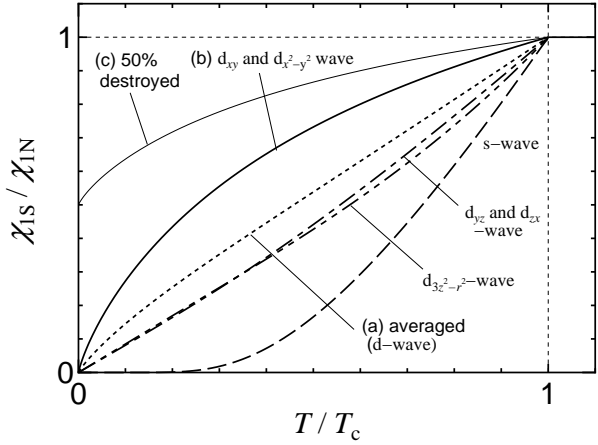}
\end{center}
\caption{
Temperature dependences of the intraband components $\chi_1$ 
of the spin susceptibilities 
for the five d-wave states and the s-wave state. 
The solid curve is 
for both ${\rm d}_{xy}$ and ${\rm d}_{x^2-y^2}$ wave pairing, 
which coincide. 
The dot-dashed curve is 
for the ${\rm d}_{yz}$ and ${\rm d}_{zx}$ wave pairing 
and 
the 2 dot-dashed curve is 
for ${\rm d}_{3z^2-y^2}$ pairing. 
The dashed curve plots the results for s-wave pairing. 
The dotted and thin solid curves present the results 
for Cases (a) and (c), respectively. 
} 
\label{fig:chi1}
\end{figure}

Figure~\ref{fig:chi1} plots the numerical results. 
The curves for the ${\rm d}_{xy}$- and ${\rm d}_{x^2-y^2}$-wave states, 
which coincide, are upward convex, while those for 
the ${\rm d}_{yz}$-, ${\rm d}_{zx}$-, ${\rm d}_{3z^2-r^2}$-, 
and s-wave states are downward convex. 
Therefore, 
the former states are more stable 
against a magnetic field along the $z$-axis 
than the latter states are.

Because of this result, 
we consider the following conditions [Cases (a) -- (c)] 
for polycrystalline powder samples: 
(a) the orientations of the powder particles are random, 
(b) for all powder particles, 
the ${\rm d}_{xy}$ or ${\rm d}_{x^2-y^2}$ wave state is realized, 
and 
(c) for a portion of the sample, 
the ${\rm d}_{xy}$ or ${\rm d}_{x^2-y^2}$ wave state is realized 
while, for the rest, the superconductivity is destroyed.

For Case (a), the gap function is oriented randomly in each powder particle. 
This can be mathematically expressed by changing the polar axis randomly 
for polar coordinates in $\Delta_{\vk +}^{(\alpha)}$, 
i.e., 
by replacing 
$\Delta_{\vk +}^{(\alpha)}(\theta_{\hat \vk}, \varphi_{\hat \vk})$ 
with 
$\Delta_{\vk +}^{(\alpha)}(\theta'_{\hat \vk}, \varphi'_{\hat \vk})$, 
where $(\theta'_{\hat \vk}, \varphi'_{\hat \vk})$ are the polar coordinates 
for the new random polar axis, 
and 
the factor $\cos^2 \theta_{\hat \vk}$ in $\chi_1$ remains unaffected. 
Since the original polar axis for $(\theta_{\hat \vk}, \varphi_{\hat \vk})$ 
is parallel to the uniform magnetic field, 
the angle between the two polar axes is random. 
Therefore, the spin susceptibility of the bulk sample is obtained 
by replacing the factor $\cos^2 \theta_{\hat \vk}$ 
with the angle average on the Fermi surface, 
which is equal to $1/3$ in the present spherically symmetric system.

Case (b) can occur 
when the system is affected by the magnetic field 
so that the spin polarization energy is lowered. 
This situation can be realized by the following mechanisms: 
(b-1) when some of the d-wave states are approximately degenerate, 
the degeneracy is lifted by the magnetic field 
in each powder particle, 
and 
(b-2) the powder particles are freely rotated by the magnetic field. 
For Case (b), 
the factor $\cos^2 {\bar \theta}_{\vk}$ in $\chi_1$ is not averaged.

Case (c) can arise 
in the presence of the impurity pair-breaking effect~\cite{Car94} 
in addition to the same conditions as Case (b). 
The anisotropic superconductivity is fragile against nonmagnetic impurities. 
As an example, we assume that the superconductivity is destroyed 
for 50\% of the powder particles.

The results for Cases (a) to (c) are shown in Fig.~\ref{fig:chi1}. 
The reduction of the spin susceptibility 
because of the growth of the superconducting gap 
in the ${\rm d}_{xy}$ and ${\rm d}_{x^2-y^2}$ wave states 
is smaller than that in the s-wave state and the other d-wave states. 
The graph for Case (a) shown in Fig.~\ref{fig:chi1} 
is the result of $\chi_1(T)$ for 
${\rm d}_{xy}$, 
${\rm d}_{yz}$, 
${\rm d}_{zx}$, and 
${\rm d}_{x^2-y^2}$ wave pairing, which coincide. 
The result for ${\rm d}_{3z^2-y^2}$ wave pairing is 
rather different from these.

Next, we compare the theoretical results in Cases (a) -- (c) 
with the experimental data given in Ref.~\citen{Nis07} 
by the following procedure: 
(i)~determine 
$\Delta K^{\rm Pd}(0) = |A_{\rm hf}| \chi_{\rm 1N}^{\rm Pd}$ 
by comparing the theoretical curve for the s-wave state 
and the experimental data of ${\rm Li_2Pd_3B}$, 
(ii)~estimate 
$\Delta K^{\rm Pt}(0) = |A_{\rm hf}| \chi_{\rm 1N}^{\rm Pt}$ 
from the value of $\Delta K^{\rm Pd}(0)$, 
(iii)~determine $K^{\rm Pt}(T_{\rm c}^{\rm Pt})$ 
from the Knight shift data above $T_{\rm c}^{\rm Pt}$ 
in ${\rm Li_2Pt_3B}$, 
and (iv)~plot theoretical curves of 
$$
     K^{\rm Pt}(T) = K^{\rm Pt}(T_{\rm c}^{\rm Pt}) 
       + \Delta K^{\rm Pt} (0) 
         \frac{\Delta \chi_1^{\rm Pt}(T)}
              {\Delta \chi_1^{\rm Pt}(0)}
     $$
in Cases (a) -- (c).

In Step (i), 
we obtain $K^{\rm Pd}(T_{\rm c}^{\rm Pd}) \approx 0.075$\% 
and $K^{\rm Pd}(0) \approx 0.0835$\%, 
which leads to $|\Delta K^{\rm Pd}(0)| \approx 0.0085$\%. 
The values of $K^{\rm Pd}(0)$ and $K^{\rm Pt}(0)$ include 
the contribution from $\chi_{\rm V} = \chi_2 + \chi_3 $. 
In Step (ii), 
considering the reduction of the density of states 
examined by Maruyama and Yanase~\cite{Mar13}, 
we assume that $\dps{\chi_{\rm 1N}^{\rm Pt} 
\sim \chi_{\rm 1N}^{\rm Pd}/2}$, 
because one of the spin-orbit split Fermi-surfaces disappears. 
Assuming that $A_{\rm hf}$'s are on the same order in the two compounds, 
we obtain $\Delta K^{\rm Pt}(0) = 0.00425$\%. 
In Step (iii), 
we obtain $K^{\rm Pt}(T_{\rm c}^{\rm Pt}) \approx 0.0725$\%, 
from the Knight shift data above $T_{\rm c}^{\rm Pt} = 2.1$~K 
in ${\rm Li_2Pt_3B}$.

\begin{figure}[htbp]
\vspace{4ex} 
\begin{center}
\includegraphics[width=8cm]{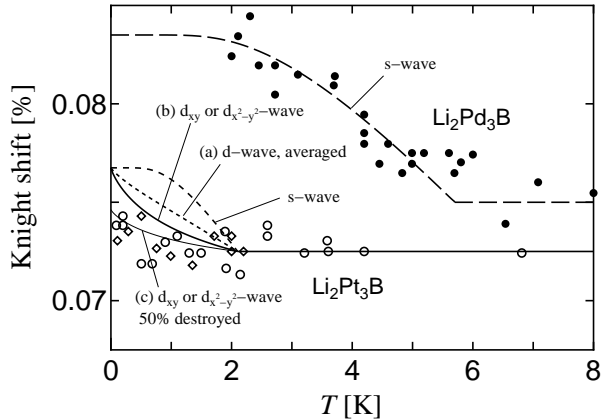}
\end{center}
\caption{
Comparison between the experimental data and the theoretical curves 
for Cases (a) -- (c). 
The closed circles, open circles, and open diamonds are 
the experimental data 
for 
${\rm Li_2Pd_3B}$ and $H = 1.46$~T, 
${\rm Li_2Pt_3B}$ and $H = 0.26$~T, 
and 
${\rm Li_2Pt_3B}$ and $H = 0.35$~T, 
respectively, 
from Ref.~\citen{Nis07}. 
The dashed and short dashed curves plot the results of s-wave pairing 
in ${\rm Li_2Pd_3B}$ and ${\rm Li_2Pt_3B}$, respectively. 
The dotted, solid, and thin sold curves present the results 
of Cases (a) -- (c) for ${\rm Li_2Pt_3B}$, respectively. 

} 
\label{fig:KnightShift}
\end{figure}

The results of Step (iv) are depicted in Fig.~\ref{fig:KnightShift}. 
For ${\rm Li_2Pt_3B}$, 
it is difficult to reproduce the experimental data 
if s-wave pairing is assumed, 
in contrast to ${\rm Li_2Pd_3B}$, 
as Nishiyama et al. have pointed out.~\cite{Nis05,Nis07} 
The theoretical result for Case (b) is in better agreement 
with the experimental data than that of Case (a). 
Since the length of the error bar shown in Ref.~\citen{Nis07} 
is approximately $\pm 0.003$\%, 
the result for Case (b) sufficiently reproduces the experimental data, 
within the experimental resolution. 
The result for Case (c) agrees very well with the experimental data, 
where it is assumed that 
the superconductivity is destroyed in half of the powder particles. 
The Knight shift slightly increases near $T = 0$ 
in the experimental data for ${\rm Li_2Pt_3B}$, 
although the increase is smaller than the error bar. 
For Cases (a) and (c), 
there are spatial distributions of spin susceptibility, 
which broaden the NMR spectra. 
However, the additional peak width would be much smaller than 
the original peak width because of dipole-dipole interaction, 
which is discussed in Ref.~\citen{Nis07}.

\section{\label{sec:summary}
Summary and Conclusions 
}

In noncentrosymmetric superconductors, 
the temperature-dependent component of the spin susceptibility $\chi_1(T)$ 
is found to exhibit an upward convex curve below $\Tc$ 
for ${\rm d}_{xy}$- and ${\rm d}_{x^2-y^2}$-wave pairing, 
and a downward convex curve 
for ${\rm d}_{yz}$, ${\rm d}_{zx}$, and ${\rm d}_{3z^2-r^2}$ wave pairing, 
when $\vH \parallel {\hat \vz}$. 
Therefore, the ${\rm d}_{xy}$ and ${\rm d}_{x^2-y^2}$ wave states are 
most stable in a magnetic field among the d-wave states. 
To explain the qualitative difference in the temperature dependence 
of the Knight shift 
in ${\rm Li_2Pd_3B}$ and ${\rm Li_2Pt_3B}$ polycrystalline powder samples, 
we have assumed three Cases (a) -- (c). 
For Case (b), the observed behaviors of the Knight shifts 
in ${\rm Li_2Pd_3B}$ and ${\rm Li_2Pt_3B}$ are consistently explained 
within the experimental resolution. 
For Case (c), the agreement between theory and experiment is excellent. 
In conclusion, the Knight shift data can be consistently explained 
if we assume that a full-gap state and a line-node state are realized 
in ${\rm Li_2Pd_3B}$ and ${\rm Li_2Pt_3B}$, respectively.

The present analysis can be extended to other anisotropic pairings 
and other forms of $\alpha_{\vk} {\vg}(\vk)$. 
If the amplitude of the gap function is large 
only in the region of $\vk$ 
where $g_z(\vk) = \cos {\bar \theta}_{\vk}$ is small, 
the temperature dependent component of the spin susceptibility $\chi_1$ 
is enhanced, 
and its temperature dependence curve can be upward convex. 
In particular, if $\Delta_{\vk} \approx 0$ for any $\vk$ 
such that $g_z(\vk) = \cos {\bar \theta}_{\vk} \ne 0$, 
we obtain $\chi_{1{\rm S}} \approx \chi_{1{\rm N}}$. 
The planar system in which $\vg(\vk)$ is perpendicular to 
a single constant vector $\vc$ for all $\vk$ 
is an extreme case. 
In this case, $\chi_{1{\rm S}} = \chi_{1{\rm N}}$ 
for any pairing anisotropy, when $\vH \parallel \vc$. 
The Rashba spin-orbit interaction ${\hat \vg}(\vk) = {\hat \vk} \times \vz$ 
is a typical example~\cite{Fri04b}. 
In such systems, 
if the orientations of the powder particles or the gap functions 
are modified by a magnetic field, as in Case (b), 
so that the spin polarization energy is minimized, 
the difference in the Knight shifts 
in the superconducting and normal phases can be small.

\mbox{}



\noindent 
{\bf Acknowledgments} 

We are very grateful to H.~Tou, T.~Shishidou, and Y.~Yanase 
for useful discussions. 
We would also like to thank G.-Q.~Zheng for the experimental data.




\end{document}